# Switching Distributions for Perpendicular Spin-Torque Devices within the Macrospin Approximation

W. H. Butler [1,2], T. Mewes[1,2], C. K. A. Mewes[1,2], P. B. Visscher[1,2], W. H. Rippard[3], S. E. Russek[3] and Ranko Heindl[3]

[1]Center for Materials for Information Technology, University of Alabama, Tuscaloosa, AL 35487 USA

[2]Department of Physics, University of Alabama, Tuscaloosa, AL 35487 USA

[3]National Institute of Standards and Technology, Boulder, CO 80305 USA

*Abstract*

We model "soft" error rates for writing (WSER) and for reading (RSER) for perpendicular spin-torque memory devices by solving the Fokker-Planck equation for the probability distribution of the angle that the free layer magnetization makes with the normal to the plane of the film.  We obtain: (1) an exact, closed form, analytical expression for the zero-temperature switching time as a function of initial angle;  (2) an approximate analytical expression for the exponential decay of the WSER as a function of the time the current is applied; (3) comparison of the approximate analytical expression for the WSER to numerical solutions of the Fokker-Planck equation; (4) an approximate analytical expression for the linear increase in RSER with current applied for reading;  (5) comparison of the approximate analytical formula for the RSER to the  numerical solution of the Fokker-Planck equation; and (6) confirmation of the accuracy of the Fokker-Planck solutions by comparison with results of direct simulation using the single-macrospin Landau-Lifshitz-Gilbert (LLG) equations with a random fluctuating field in the short-time regime for which the latter is practical.

**I. Introduction**

Spin-polarized electrical currents can transfer angular momentum between nanometer scale ferromagnetic electrodes separated by a non-magnetic layer [1,2]. The use of this effect to switch the direction of magnetization of a ferromagnetic layer as part of a magnetic memory device is of great current interest. In order to build a memory device one needs at least two distinct physical states that one can associate with the two logic states. In addition, one must have a means of switching the device between these states and a means of determining its state. A memory device is useful only if it switches (with very high probability) when switching is intended and does not switch (again with very high probability) when switching is not intended. In the most common type of spin-torque memory device, the two states are provided by the relative orientations of the direction of magnetization of two ferromagnetic layers, typically parallel and anti-parallel. The switching is achieved by the transfer of angular momentum carried by spin-polarized current, and the change in resistance between the parallel and anti-parallel states is used to determine the state of the memory device.

In this paper, we consider spin-torque devices in which the magnetization of the two ferromagnetic layers is oriented perpendicular to the film plane in the quiescent state, *i.e.* in the absence of applied field or current. One layer is considered to have fixed magnetization, whereas the other layer's magnetization is free to precess and switch. Such devices should switch at lower currents than devices (with equivalent thermal stability) for which the magnetization is in the plane of the layers in the quiescent state [3]. At the present time, most of the effort aimed at practical spin-torque memory devices uses magnetic tunnel junctions because of the high spin-polarization that can be achieved through the symmetry-based spin-filter effect [4-7] and because they offer the possibility of good impedance matching with a transistor that is used to select a particular device for reading or writing.

A desirable memory device should switch both quickly and reliably when switching is intended and it should not switch when switching is not intended, for example when current is applied to read the state of the device. Our concern will be the probability of switching events when switching is not intended (read soft-error rate or RSER), or non-switching events when switching is intended (write soft-error rate or WSER).

A very large and rapidly growing literature on spin-torque switching, from the perspectives of both theory[1,2,8-25] and experiment[25-29], now exists. In the following we will focus on the following aspects of the spin-torque switching phenomenon: (1) Additional terms in the LLG equation that arise from Gilbert damping acting on the spin-torque; (2) exact integration of the zero temperature switching equation; (3) numerical solution to the Fokker-Planck equation giving the finite temperature distribution of switching elements; (4) approximate analytic solution to the Fokker-Planck equation valid for currents above the critical current for switching; (5) approximate analytic solution for the switching rate for currents below the critical current for switching; (6) demonstration that, for the purpose of determining switching distributions, the Fokker-Planck equation is equivalent to macrospin simulations that include a random thermal field, with the exception that the Fokker-Planck approach can be applied to determine switching probabilities that are extremely small or very close to unity, thus allowing the investigation of WSER and RSER; and (7) observation that, for the presently considered case with axial symmetry, the equations controlling switching via spin-polarized currents can be mapped onto mathematically equivalent equations for switching via an applied axial magnetic field.

**Landau-Lifshitz-Gilbert (LLG) Equation for Film with Perpendicular Anisotropy**

Consider the evolution in time of the magnetization angular momentum, **L**, of the free layer (treated here as a macrospin, meaning that internal magnetic degrees of freedom are ignored) of a magnetic tunnel junction which is receiving electrons through a tunnel barrier from a pinned layer that has its magnetization pinned in the direction, $\hat{\mathbf{m}}_p$. The rate of change of **L** can be written (in SI units) as [2,30,31]

$$\frac{\partial \mathbf{L}}{\partial t} = -\mu_0 V \mathbf{M} \times \mathbf{H}_{\text{eff}} + \alpha \hat{\mathbf{m}} \times \frac{\partial \mathbf{L}}{\partial t} - \eta \frac{\hbar}{2} \frac{I}{e} \hat{\mathbf{m}} \times (\hat{\mathbf{m}} \times \hat{\mathbf{m}}_p) + \frac{\hbar}{2e} f(I)(\hat{\mathbf{m}} \times \hat{\mathbf{m}}_p), \qquad (1.1)$$

where the first term on the right hand side of Eq. (1.1) is the torque on the magnetic moment of the free layer, $V\mathbf{M}$ ($V$ is the volume of the free layer and $\mathbf{M} = M\hat{\mathbf{m}}$ is its magnetization), due to the effective field, $\mathbf{H}_{\text{eff}}$; the second term is an empirical damping term designed to damp the precession of **L**; and the third and fourth terms are spin-torque terms that arise from spin angular momentum carried by the electrons that tunnel through the barrier carrying angular momentum from the pinned layer to the free layer or vice versa. Each of these terms will be briefly discussed in turn.

The magnetic moment and angular momentum are related through the gyromagnetic ratio, $\gamma$, by, $\mathbf{L} = \mathbf{M}V/\gamma$, so the first term causes the angular momentum to precess in a direction perpendicular to both the effective field and the angular momentum. The effective field arises from the anisotropy, demagnetizing effects, and any external field. In this paper we shall assume that all three of these contributions are perpendicular to the plane of the film. This assumption significantly simplifies the analysis. The additional assumption that $\hat{\mathbf{m}}_p$ is also perpendicular to the plane of the film is sufficient to provide the problem with axial symmetry. To avoid ambiguity, we note that the gyromagnetic ratio, $\gamma$, as used in this paper is in units of *angular frequency per Tesla*, *i.e.* $\gamma \approx 1.76 \times 10^{11}$ radians/(Ts).

The second term is perpendicular to both the angular momentum and its precessional motion. This term drains energy from the precessing free layer moment and eventually causes the moment and angular momentum vector to align with $\mathbf{H}_{eff}$. Note that any term that causes torque (i.e., the third and fourth terms as well as the first) is expected to contribute to the damping because the semi-empirical Gilbert damping term is proportional to *dL/dt*, regardless of its origin.

The torque described by the third term can be understood heuristically as follows: *I*/e is the number per second of spin-polarized electrons entering the free layer when moments of the free and pinned layers are perpendicular. Each of these electrons carries angular momentum, $\hbar/2$. The parameter $\eta$ is given in terms of the currents for collinear orientation of the moments of the pinned and free layers by [10,14,17]

$$\eta = \frac{I_{++} + I_{+-} - I_{--} - I_{-+}}{I_{++} + I_{+-} + I_{--} + I_{-+}}. \tag{1.2}$$

Here $I_{++}$ and $I_{--}$ are the majority and minority currents respectively that would flow for parallel alignment of the layers, while $I_{+-}$ and $I_{-+}$ are the majority and minority currents (from the perspective of the pinned layer) respectively for anti-parallel alignment of the two layers. The angle-dependent term, $\hat{\mathbf{m}} \times (\hat{\mathbf{m}} \times \hat{\mathbf{m}}_p)$ accounts for the fact that only the transverse part of the incoming angular momentum can be absorbed by the free layer. In a spherical coordinate system in which the polar angle is measured from the normal to the plane and $\hat{\mathbf{m}}_p$ is also in this direction, this torque is in a direction to increase or decrease the polar angle. For brevity and convenience we refer to this term as the Slonczewski spin-torque term.

The torque described by the fourth term is often described as the "field-like" spin-torque and cannot be simply expressed in terms of the electron particle current [17,18]. The nomenclature "field-like" is based on the fact that as long as $f(I)$ does not depend on the angle between $\hat{\mathbf{m}}$ and $\hat{\mathbf{m}}_p$ this term can be treated as an additional contribution to the effective field, primarily increasing or decreasing the rate of precession, but also changing the effective energy function, i.e. the effective energy as a function of polar angle that determines the switching rate.

The LLG equation (1.1) can be written in the classical Landau-Lifshitz form by converting to an equation for $\mathbf{M}$ using $\mathbf{L}=\mathbf{M}V/\gamma$, and then inserting the terms on the right-hand side of the equation for $\frac{\partial \mathbf{M}}{\partial t}$ and simplifying,

$$(1+\alpha^2)\frac{\partial \mathbf{M}}{\partial t} = -\gamma\mu_0 M_S \hat{\mathbf{m}} \times \mathbf{H}_{eff} - \frac{\gamma\eta I\hbar}{2eV}\hat{\mathbf{m}} \times (\hat{\mathbf{m}} \times \hat{\mathbf{m}}_p) + \frac{\gamma\hbar f(I)}{2eV}(\hat{\mathbf{m}} \times \hat{\mathbf{m}}_p)$$
$$-\alpha\gamma\mu_0 M_S \hat{\mathbf{m}} \times \hat{\mathbf{m}} \times \mathbf{H}_{eff} + \frac{\alpha\gamma\eta I\hbar}{2eV}(\hat{\mathbf{m}} \times \hat{\mathbf{m}}_p) + \frac{\alpha\gamma\hbar f(I)}{2eV}\hat{\mathbf{m}} \times (\hat{\mathbf{m}} \times \hat{\mathbf{m}}_p).$$
(1.3)

Written this way, we can see that the effect of the Gilbert damping term, $\alpha\hat{\mathbf{m}} \times \frac{d\mathbf{L}}{dt}$, is to renormalize the rate of precession by a factor $1+\alpha^2$ (e.g., by effectively reducing the gyromagnetic ratio) and to add a term, $-\alpha\gamma\mu_0 M_S \hat{\mathbf{m}} \times \hat{\mathbf{m}} \times \mathbf{H}_{eff}$, which yields a contribution to $\frac{\partial \mathbf{M}}{\partial t}$ in the direction of the polar angle that acts to return the magnetization towards $\mathbf{H}_{eff}$. In addition to these two well known effects, the Gilbert damping also acts on the spin-torque terms so that the Slonczewski spin-torque term generates a field-like torque and the field-like torque term generates a torque in the same direction as the Slonczewski term. Both of these terms are reduced by a factor of the Gilbert damping parameter, $\alpha$.

We obtain the effective field, $\mathbf{H}_{eff}$, using

$$\mu_0 \mathbf{H}_{eff} = -\left(\frac{\partial E}{\partial \mathbf{M}}\right) = -\left(\hat{x}\frac{\partial E}{\partial M_x} + \hat{y}\frac{\partial E}{\partial M_y} + \hat{z}\frac{\partial E}{\partial M_z}\right),$$
(1.4)

where $E$ is the magnetic energy per unit volume of the free layer. If the free layer has no in-plane anisotropy, $E$ can be written as,

$$E = \tfrac{1}{2}\mu_0 N_{zz} M_z^2 + \left(K_U^B + \frac{K_U^S}{t}\right)\left(1 - \frac{M_z^2}{M_s^2}\right) - \mu_0 M_z H_{z\text{-ext}},$$
(1.5)

where the first and second terms come from demagnetization and anisotropy, respectively. $N_{zz}$ (<1) is the $zz$ component of the demagnetization tensor. The anisotropy may arise from bulk magnetocrystalline anisotropy ($K_U^B$) or from surface anisotropy ($K_U^S/t$). Here, $t$ represents the thickness of the magnetic free layer. The third term comes from an external applied magnetic field. In this paper, in order to preserve axial symmetry, we restrict our treatment to the case in which any external field is perpendicular to the layers. In fact, our primary interest is in RSER and WSER with no external field present. The external field is included here partly for completeness, but more importantly, because it allows us to make a connection with previous work by Brown [32-34] and others [35-39].

In the macrospin model, the magnitude of the magnetic moment of the free layer is assumed to be constant in time. Thus, the important quantity is its direction, which we will describe in spherical polar coordinates, *i.e.,* $\hat{\mathbf{m}} = \sin\theta(\cos\phi\hat{\mathbf{x}} + \sin\phi\hat{\mathbf{y}}) + \cos\theta\hat{\mathbf{z}}$. The magnetic energy per unit volume (relative to its value for $\theta = 0$), expressed in spherical polar coordinates is,

$$E = K_U^{\text{eff}} \sin^2\theta - \mu_0 M_S H_{\text{z-ext}} (\cos\theta - 1), \tag{1.6}$$

where $K_U^{\text{eff}} = K_U^B + K_U^S/t - \tfrac{1}{2}\mu_0 N_{zz} M_S^2$. This energy expression leads to an effective magnetic field,

$$\mathbf{H}_{\text{eff}} = \left(H_K^{\text{eff}} \cos\theta + H_{z-\text{ext}}\right)\hat{z} = H_K^{\text{eff}}(\cos\theta + h)\hat{\mathbf{z}}, \tag{1.7}$$

where $H_K^{\text{eff}} = \dfrac{2K_U^{\text{eff}}}{\mu_0 M_s}$ and $h = H_{z-\text{ext}} / H_K^{\text{eff}}$. $H_K^{\text{eff}}$ is called the switching field because a field of this magnitude applied along the easy axis (perpendicular to the layers) will cause the free layer to switch. In the following, we shall suppress the superscript on $H_K$. It will be understood that $H_K$ is the effective anisotropy field resulting from both anisotropy and demagnetization.

**Equations of Motion**

Substitution of equation (1.7) for $\mathbf{H}_{\text{eff}}$ in equation (1.3) yields,

$$(1+\alpha^2)\frac{\partial \hat{\mathbf{m}}}{\partial t} = -\gamma\mu_0 H_K \left[\cos\theta + h + \alpha^2 \bar{i} + \frac{\alpha}{\eta}\frac{f(I)}{I_0}\right]\hat{\mathbf{m}}\times\hat{\mathbf{z}}$$
$$- \alpha\gamma\mu_0 H_K \left[\cos\theta + h - \bar{i} + \frac{\alpha}{\eta}\frac{f(I)}{I_0}\right]\hat{\mathbf{m}}\times(\hat{\mathbf{m}}\times\hat{\mathbf{z}}). \quad (2.1)$$

Here, we have introduced a reduced current, $\bar{i} = \frac{I}{I_0}$, where $I_0$ is defined by

$$I_0 = \frac{\alpha}{\eta}\frac{2\text{e}}{\hbar}\mu_0 H_K M_S V, \quad (2.2)$$

and is the critical current for switching via the first spin-torque term. The terms in equation (2.1) proportional to $\hat{\mathbf{m}}\times\hat{\mathbf{z}}$ yield a torque in the azimuthal ($\phi$) direction and contribute to precession around the $\hat{\mathbf{z}}$ axis. On the other hand, the terms proportional to $\hat{\mathbf{m}}\times(\hat{\mathbf{m}}\times\hat{\mathbf{z}})$, generate a torque that changes the polar angle and are responsible for spin-torque switching.

In terms of $\theta$ and $\phi$, the Landau-Lifshitz-Gilbert equation becomes

$$(1+\alpha^2)\frac{d\phi}{dt} = \omega_0\left(\cos\theta + h + \alpha^2\bar{i} + \frac{\alpha}{\eta}\frac{f(I)}{I_0}\right), \quad (2.3)$$

$$(1+\alpha^2)\frac{d\theta}{dt} = \alpha\omega_0 \sin\theta\left(\bar{i} - h - \frac{\alpha}{\eta}\frac{f(I)}{I_0} - \cos\theta\right), \quad (2.4)$$

where $\omega_0 = \gamma\mu_0 H_K$. Equation (2.3) for $\phi$ simply describes the precession of the magnetic moment around the $\hat{\mathbf{z}}$ axis. We will not consider it further except to note that the field-like term (and to a lesser extent the Sloncewski-torque, since typically $\alpha<<1$) yields a current or bias dependence to the FMR frequency that might be observable. Equation (2.4) for $\theta$ describes the variation of the polar angle as a competition among the Slonczeweski spin-torque term (proportional to $\bar{i}$), field (proportional to $h$), and the damping term (proportional to $\cos\theta$). Depending on their signs, the first two either increase or decrease $\theta$, whereas the third always decreases it (if $\theta < \pi/2$). Note that the field-like term can be viewed either as a modification to $i$ or a modification to $h$. Physically, modification of $h$ may seem more appealing since this term originates from the change in damping due to the change in the rate of precession, similar to that caused by an applied field. Practically, however, it is more convenient to include the field-like term with the Slonczewski term because both are controlled by the electrical current or bias.

We can simplify our equation even more if we measure time in units of $(1+\alpha^2)/(\alpha\omega_0)$. Thus, if our unit of time is that required for $(1+\alpha^2)/(2\pi\alpha)$ precessional orbits (in the absence of applied fields or currents) we have

$$\frac{d\theta}{d\tau} = (i - h - \cos\theta)\sin\theta, \tag{2.5}$$

where

$$\tau = \frac{\alpha\gamma\mu_0 H_K}{1+\alpha^2} t, \tag{2.6}$$

and

$$i = \left(1 - \frac{\alpha}{\eta}\frac{f(I)}{I}\right)\frac{I}{\frac{\alpha}{\eta}\frac{2e}{\hbar}\mu_0 H_K M_S V}. \tag{2.7}$$

Since $f(I)/I$ may not be independent of $I$, $i$ may not be precisely proportional to $I$. We shall nevertheless, for simplicity, refer to $i$ as the "reduced current".

The critical current for switching is an important quantity for spin-torque devices because the magnitude of the current that must be supplied determines the size of the transistor needed to supply the current and ultimately the energy required for switching. Another important quantity is the thermal stability factor, $\Delta$, defined as the energy barrier for switching divided by the thermal energy, $k_BT$,

$$\Delta = \frac{K_U^{\text{eff}} V}{k_B T} = \frac{\mu_0 H_K M_S V}{2 k_B T}. \tag{2.8}$$

The thermal stability factor determines the rate at which thermal fluctuations cause an element to switch. To avoid thermally induced switching (over practically relevant periods), $\Delta$ should be greater than ~50. The critical current for switching is proportional to the thermal stability factor,

$$I_0 = \frac{\alpha}{\eta} \frac{2e}{\hbar} 2\Delta k_B T \approx \frac{\alpha}{\eta} 1.5 \text{mA}. \tag{2.9}$$

In the limit of $\theta \to 0$, equation (2.5) becomes

$$\frac{d\theta}{d\tau} = (i - h - 1)\theta, \tag{2.10}$$

with solution,

$$\theta = \theta_0 \exp[(i - h - 1)\tau]. \tag{2.11}$$

The switching time in this approximation is [40]

$$\tau_{sw} = \frac{\ln\left(\frac{\pi}{2\theta_0}\right)}{i - h - 1}. \tag{2.12}$$

This expression is based on the notion that the element switches when $\theta$ equals $\pi/2$.

Equation (2.5) can also be solved without making the $\theta \to 0$ approximation. Substituting $x = \cos\theta$ into equation (2.5), we have,

$$\frac{dx}{d\tau} = -(i-h-x)(1-x^2), \qquad (2.13)$$

which can be integrated to give

$$\tau(x) = \frac{\frac{1}{2}(i-h+1)\ln\left(\frac{1-x}{1-x_0}\right) + \frac{1}{2}(i-h-1)\ln\left(\frac{1+x_0}{1+x}\right) + \ln\left(\frac{i-h-x_0}{i-h-x}\right)}{(i-h)^2 - 1}, \qquad (2.14)$$

where $x_0$ is the value of $\cos\theta$ at $\tau=0$. When $\theta$ reaches $\pi/2$, $\cos\theta$ changes sign so that the damping term changes from impeding switching by pushing $\theta$ back towards zero to assisting switching by pushing $\theta$ towards $\pi$, thus it is reasonable to assume that the element switches when $\theta=\pi/2$ or $\cos\theta=0$, yielding,

$$\tau_{sw} = \frac{-\frac{1}{2}(i-h+1)\ln(1-x_0) + \frac{1}{2}(i-h-1)\ln(1+x_0) + \ln\left(1 - \frac{x_0}{i-h}\right)}{(i-h)^2 - 1}. \qquad (2.15)$$

**Fokker-Planck Equation**

Equation (2.15) gives the time required for an element that has its moment pointing at an angle $\theta_0 = \cos^{-1}x_0$ at $\tau=0$ to switch. It implicitly assumes that the motion is deterministic, *i.e.*, that neither the initial displacement angle, $\theta_0$, nor the trajectory in $\theta, \phi$ space are affected by thermal fluctuations. In actuality, of course, thermal fluctuations will have important effects upon the switching. We are primarily interested in the probability of rare events, namely switching that occurs when not intended (when current is applied for reading the state of the system) or switching that does *not* occur when switching is intended (when current is applied for switching). One approach to this problem is to solve the Landau-Lifshitz-Slonczewski equation (Eq.(1.1)) with an additional, random thermal field many times while recording whether the magnetization has switched. We will use this approach to validate our results based on the Fokker-Planck method described in the following.

We derive and solve a Fokker-Planck equation [32,41,42] for the probability distribution of the angle, $\theta$, as a function of $\tau$. We define $\rho(\theta,\tau)$ to be the probability that the magnetization is pointing in direction $\theta$ relative to the film normal at time $\tau$. $\rho(\theta,\tau)$ is normalized so that

$$\int_0^\pi \rho(\theta,\tau)\sin\theta d\theta = 1 \tag{3.1}$$

for all $\tau$. The Fokker-Planck equation is based on the continuity equation for this probability density,

$$\frac{\partial \rho(\theta,\tau)}{\partial \tau} = -\nabla \cdot \mathbf{J}(\theta,\tau) = -\frac{1}{\sin\theta}\frac{\partial(\sin\theta J_\theta(\theta,\tau))}{\partial \theta}, \tag{3.2}$$

which simply states that the rate of change of the probability density at angle $\theta$ is equal to the net rate at which probability density flows in. The current in probability density consists of a flow term and a diffusion term,

$$J_\theta(\theta,\tau) = \rho(\theta,\tau)\frac{\partial \theta}{\partial \tau} - D\frac{\partial \rho(\theta,\tau)}{\partial \theta}, \tag{3.3}$$

where $D$ is the diffusion coefficient and $\frac{\partial \theta}{\partial \tau}$ is given by equation (2.10). Combining equations (3.2) and (3.3) yields

$$\frac{\partial \rho(\theta,\tau)}{\partial \tau} = -\frac{1}{\sin\theta}\frac{\partial}{\partial \theta}\left(\sin^2\theta(i-h-\cos\theta)\rho(\theta,\tau) - D\sin\theta\frac{\partial \rho(\theta,\tau)}{\partial \theta}\right). \tag{3.4}$$

The diffusion constant, $D$, can be determined from the equilibrium condition in the absence of an applied electrical current or field, because in this case equation (3.4) reduces to

$$\frac{\partial \rho_{eq}(\theta)}{\partial \theta} = \frac{-1}{D}\sin\theta\cos\theta\rho_{eq}(\theta), \tag{3.5}$$

which has the solution

$$\rho_{eq}(\theta) = A\exp\left(\frac{-\sin^2\theta}{2D}\right). \tag{3.6}$$

On the other hand, the equilibrium distribution should be Maxwell-Boltzmann, which implies

$$\rho_{eq}(\theta) \propto \exp\left[-\frac{E(\theta)}{k_B T}\right] = \exp\left[-\Delta\sin^2\theta\right]. \tag{3.7}$$

Thus, $D = \frac{1}{2\Delta}$, so the Fokker-Planck equation is

$$\frac{\partial \rho(\theta,\tau)}{\partial \tau} = -\frac{1}{\sin\theta}\frac{\partial}{\partial \theta}\left(\sin^2\theta(i-h-\cos\theta)\rho(\theta,\tau) - \frac{\sin\theta}{2\Delta}\frac{\partial \rho(\theta,\tau)}{\partial \theta}\right). \tag{3.8}$$

It can also be expressed in terms of $x = \hat{m}_z = \cos\theta$ as

$$\frac{\partial \rho(x,\tau)}{\partial \tau} = \frac{\partial}{\partial x}\left[(i-h-x)(1-x^2)\rho + \frac{(1-x^2)}{2\Delta}\frac{\partial \rho}{\partial x}\right]. \tag{3.9}$$

Equation (3.9) is equivalent, aside from the term *i* in *(i - h - x)* to an expression derived by Brown [32] for the similar reversal of a macrospin by a magnetic field.

**Approximate Analytical Solutions**

We can obtain an approximate analytical solution to equation (3.8) in the limit in which $\theta$ is small. In this limit, (3.8) becomes,

$$\frac{\partial \rho(\theta,\tau)}{\partial \tau} \approx -\frac{1}{\theta}\frac{\partial}{\partial \theta}\left(\theta^2(i-h-1)\rho(\theta,\tau) - \frac{\theta}{2\Delta}\frac{\partial \rho(\theta,\tau)}{\partial \theta}\right). \tag{3.10}$$

To solve this equation, we will use the ansatz, $\rho(\theta,\tau) = A(\tau)\exp\left(-\dfrac{\theta^2}{\bar{\theta}^2(\tau)}\right)$. $A(\tau)$ and $\bar{\theta}^2(\tau)$ can be related through the normalization condition,

$$1 = \int_0^\pi \sin\theta\, d\theta\, \rho(\theta,\tau) = \int_0^\pi \sin\theta\, d\theta\, A(\tau)\exp\left(-\dfrac{\theta^2}{\bar{\theta}^2(\tau)}\right) \approx \dfrac{1}{2} A(\tau)\bar{\theta}^2(\tau), \qquad (3.11)$$

which implies that $A(\tau) = \dfrac{2}{\bar{\theta}^2(\tau)}$. Substitution of this ansatz into equation (3.10) yields the following equation for $\bar{\theta}^2(\tau)$:

$$\dfrac{\partial \bar{\theta}^2(\tau)}{\partial \tau} = 2(i-h-1)\bar{\theta}^2(\tau) + \dfrac{2}{\Delta}. \qquad (3.12)$$

The solution to equation (3.12) subject to the boundary condition, $\bar{\theta}^2(\tau) \to \dfrac{1}{\Delta}$ for $\tau \to 0$, is

$$\bar{\theta}^2(\tau) = \dfrac{(i-h)\exp(2\tau(i-h-1)) - 1}{\Delta(i-h-1)}, \qquad (3.13)$$

so that the distribution function is approximated as,

$$\rho(\theta,\tau) = \dfrac{2}{\bar{\theta}^2(\tau)}\exp\left(-\dfrac{\theta^2}{\bar{\theta}^2(\tau)}\right) \qquad (3.14)$$

with $\bar{\theta}^2(\tau)$ given by Eq. (3.13).

Note that equation (3.13) predicts that at large times ($2\tau(i-h-1) \gg 1$) the distribution will become independent of $\theta$ and decay as $\rho \to \dfrac{2\Delta(i-h-1)}{i-h}\exp[-2\tau(i-h-1)]$. The switched fraction in this approximation, calculated by assuming that an element switches at $\theta = \pi/2$, is

$$P_{sw} = \exp\left(-\dfrac{\pi^2}{4\bar{\theta}^2(\tau)}\right). \qquad (3.15)$$

The accuracy of this approximation is tested in the next section by comparison to numerical solutions of the Fokker-Planck equation that do not invoke the small-angle approximation.

**Numerical Solutions**

Equation (3.9) for $h=0$, is solved on the line $-1 \leq x \leq 1$, where $x = \cos\theta$. The usual starting distribution is,

$$\rho_{initial}(\theta) = \frac{\exp(-\Delta \sin^2\theta)\Theta\left(\frac{\pi}{2}-\theta\right)}{\int_0^{\pi/2} \sin\theta \exp(-\Delta \sin^2\theta) d\theta} \tag{4.1}$$

or

$$\rho_{initial}(x) = \frac{\exp(-\Delta(1-x^2))\Theta(x)}{\int_0^1 \exp(-\Delta(1-x^2)) dx} \tag{4.2}$$

where the Heaviside function, $\Theta\left(\frac{\pi}{2}-\theta\right)$, constrains the initial distribution to the well near $\theta = 0$. The calculated probability distribution for $\Delta=60$ and $i=1.5$ is shown in Figure 1. Initially, the probability is confined to the well at $\theta = 0$. By reduced time $\tau \sim 4$, however, the distribution on a semi-log plot takes on a shape that does not vary qualitatively with time. It is a relatively uniform distribution over $\theta$, except for a concentration near $\theta=\pi$.

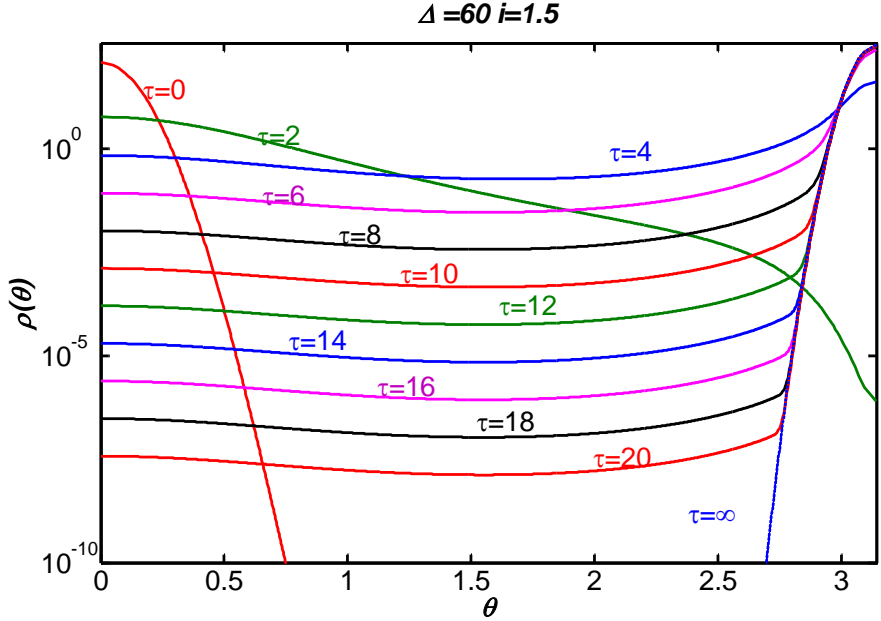

**Figure 1. (Color online) Probability distribution for different reduced times, (numerical solution to equation (3.8) or (3.9) for $\Delta=60$ and $i=1.5$.**

As time increases, the approximately uniform part decreases exponentially as the probability flows into the concentration near $\theta=\pi$. At long times, the distribution function becomes

$$\rho(\theta,\infty) = N\,e^{-\Delta\left(\sin^2\theta + 2i\cos\theta\right)}, \tag{4.3}$$

where $N$ is a normalization factor. This function gives the bell shaped curve labeled $\tau=\infty$ on the right side of Figure 1. It provides a particular solution to equation (3.8) for which $\dfrac{\partial\rho(\theta,\tau)}{\partial\tau}=0$. Equation (4.3) is easily understood from equation (2.5), which implies that the spin-torque current, insofar as it enters the equation for $d\theta/d\tau$, acts like an additional axial magnetic field. Thus the Fokker-Planck equation for perpendicular spin-torque systems is equivalent to the one dimensional diffusion equation for a particle in a potential of form,

$$E(\theta)V = K_U V\left[\sin^2\theta + 2(i-h)(\cos\theta-1)\right]. \tag{4.4}$$

The $\tau \to \infty$ solution, Eq. (4.3), is the spin-torque analog of an expression derived by Brown [32] for the steady state solution to the Fokker-Planck equation for a Stoner-Wohlfarth element in an easy axis field.

Figure 2 compares the numerical solution shown in Figure 1 with the approximate solution, equation (3.13), which certainly is not valid in the well centered at $\theta = \pi$, since it is based on an expansion around the bottom of the well at $\theta = 0$. For this reason, we show in Figure 2 only the part of the distribution between $\theta=0$ and $\theta=\pi/2$. The fraction of $\rho$ that has not switched can be obtained by integrating over this part of the probability distribution function.

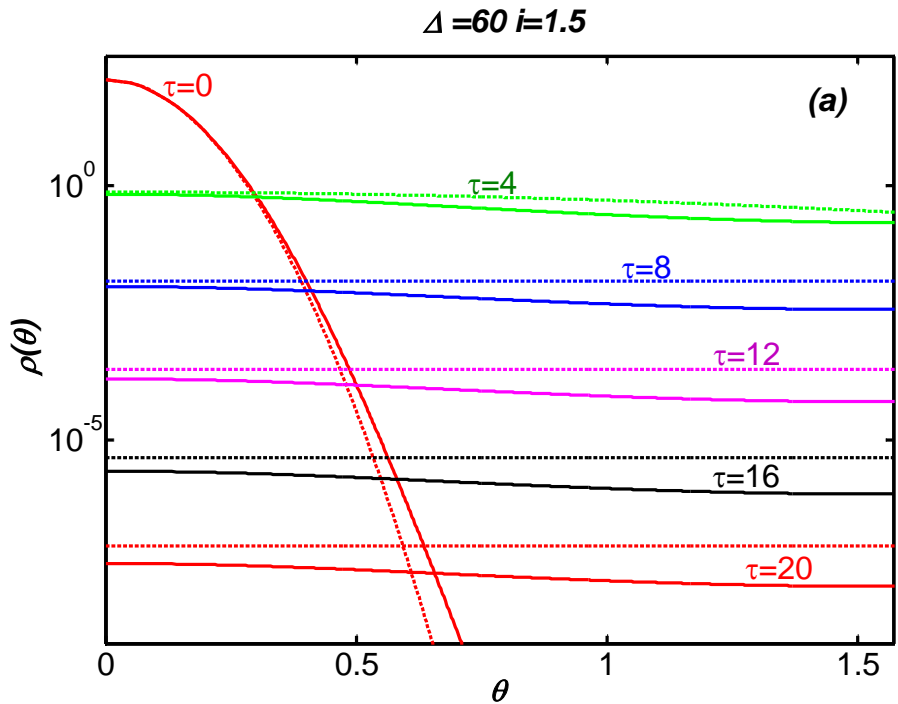

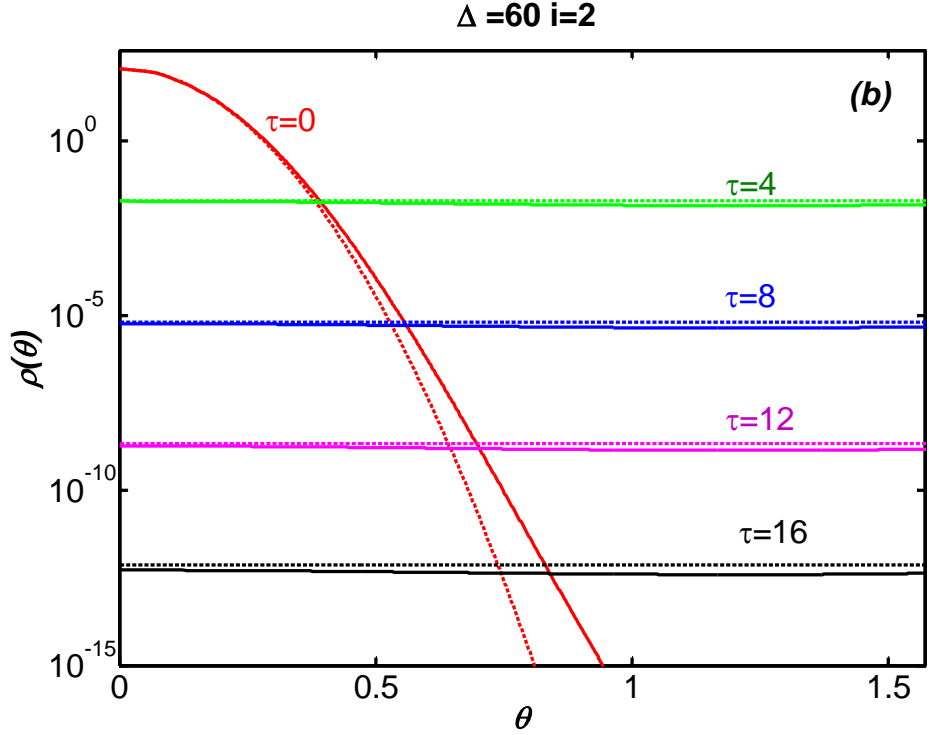

Figure 2. (Color online) Probability distribution within the well centered at $\theta=0$ for reduced currents (see Eq. (2.7)) $i=1.5$ (a) and for $i=2$ (b). Dashed lines are the approximate solution of Eq. (3.13). For both calculations, the thermal stability factor is $\Delta=60$. The reduced time (see Eq.(2.6)) is labeled by $\tau$.

The approximate analytical solution becomes more accurate as the reduced current, $i$, is increased, as can be seen by comparing figures 2(a) and 2(b). Note that the approximate solution consistently overestimates the probability distribution in the well at $\theta=0$. This is understandable because the actual equation of motion for $\theta$, $\frac{d\theta}{d\tau} = (i - h - \cos\theta)\sin\theta$, is approximated by setting $\cos\theta$ to 1, which significantly underestimates the relative amount by which the spin-torque term exceeds the damping term, especially when $i \approx 1$. Thus the frequency of write errors is less than would be predicted by the approximate analytical model, especially for low values of $i$-1.

**Effect of Thermal Stability Factor**

Figure 3 shows the time evolution of the distribution function for a thermal stability factor $\Delta=30$, one half the value used in the comparable calculation shown in Figure 1. The initial and final distribution functions are noticeably broader as would be expected. Somewhat surprisingly, the effect of this large reduction in $\Delta$ on the distribution functions at intermediate times is relatively modest. There is a relatively small (on this exponential scale) shift downward in the distribution function. The spacing of the curves for different times hardly changes. The implication of this result is that soft error rates for writing associated with non-switching events should be relatively insensitive to the thermal stability factor.

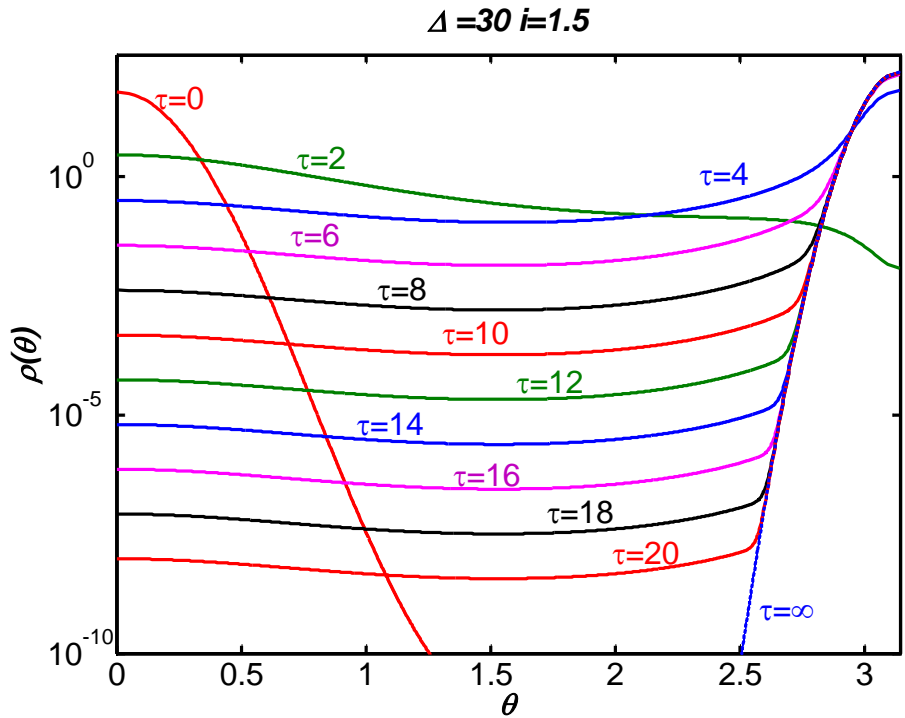

**Figure 3. Distribution functions for thermal stability factor, $\Delta=30$ for reduced time ranging from 0 to 20. The time is measured in units of the FMR period divided by the damping parameter (see Eq. (2.6)).**

Figure 4a shows the soft error rate for writing (WSER) as a function of time for 5 values of $i$ with $\Delta$ fixed at 60 (solid lines). Similar results are presented for $\Delta=30$ in Figure 4b. For comparison, we also show in both figures the WSER calculated from $(1-P_{sw})$ where $P_{sw}$ is given by the analytic approximation (3.15). The analytic approximation overestimates $P_{ns}$, especially for low values of $i-1$, but accurately represents the exponential decay which is well described by $\exp[-2\tau(i-1)]$. It can be seen that $\Delta$ has only a small effect on the WSER.

These results place limits on the write speed of perpendicular spin-torque memory elements based on the simple structures considered here and for which the macrospin approximation is applicable. In Figure 4, time is measured in dimensionless units $\tau = \alpha\gamma\mu_0 H_K t/(1+\alpha^2) = t/t_0$ where $t_0 = T/(2\pi\alpha)$ and $T$ is the ferromagnetic resonance period (for no applied field or current). Thus, if the damping parameter, $\alpha$, is 0.01 and a WSER of $10^{-9}$ is desired, Figure 4 predicts that a current of $2I_0$ should be applied for approximately $1200/(2\pi)$ FMR periods. If the applied current is $1.5\ I_0$, the current should be applied for nearly twice that time. If the FMR frequency is 10 GHz, $1200/(2\pi)$ FMR periods requires 19.1 ns. These results also point to obvious design criteria that can be used to reduce the switching time while maintaining the desired WSER.

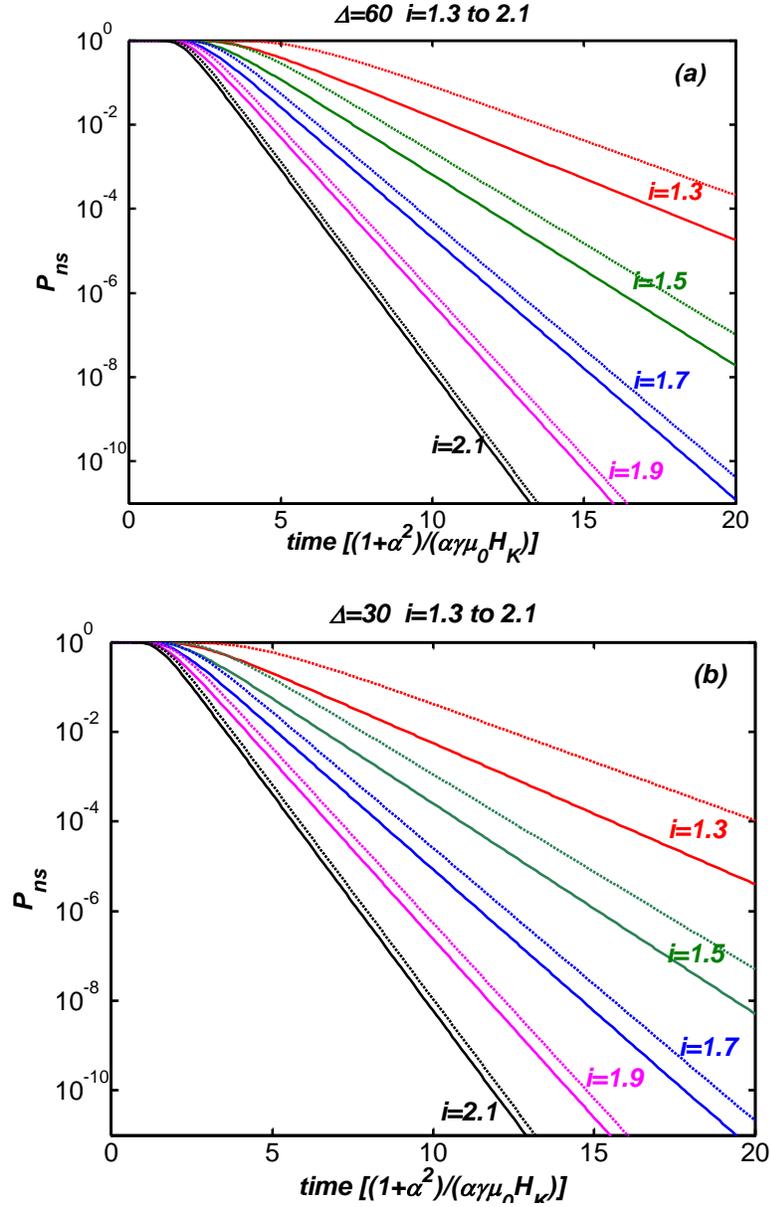

**Figure 4.** Non-switched probability, soft-error rate for writing using a thermal stability factor of (a) $\Delta=60$ and (b) $\Delta=30$ as a function of time for several values of the reduced current. Current, *i*, is measured in units of $I_0$, the critical current for switching and time is measured in units of the FMR period divided by $2\pi$ times the damping parameter. Solid lines are solutions of the Fokker-Planck equation. Dashed lines represent the approximate analytic solution (see Eq.(3.15)).

In figure 5 we show a comparison between the numerical solution of the Fokker-Planck equation and macrospin simulations including a thermal field [43]. Both calculations were carried out for $\Delta=60$ and for $i=2,3,4,5$ and 6. The macrospin simulations were done by running each simulation for a fixed time interval and determining the state (unswitched or switched) at each simulation time step, thereby taking into account the possibility of rare switch-back events. This approach also leads to apparent correlations between $P_{ns}(t)$ and $P_{ns}(t+\Delta t)$.

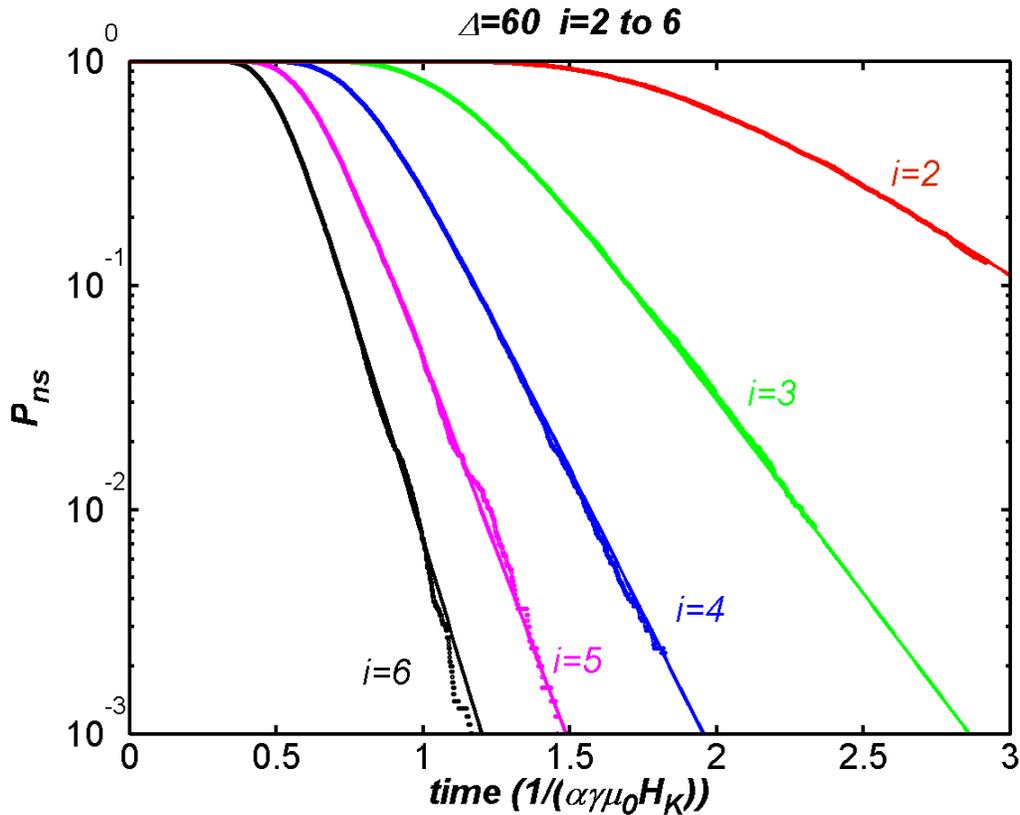

**Figure 5 . Non-switched probability, soft-error rate for writing using a thermal stability factor of $\Delta=60$ as a function of time for several values of the reduced current $i$ measured in units of $I_0$, the critical current for switching. Time is measured in units of the FMR period divided by the $2\pi$ times the damping parameter. Smooth solid curves are solutions of the Fokker-Planck equation. Jagged lines composed of discrete points are result from Landau-Lifshitz simulations of a macrospin including a thermal field[43] as described in the text.**

Both the Fokker-Planck calculations and the simulations were based on the LL rather than the LLG formulation for damping and both omitted the "field-like" term. (Both of these omissions can be included by a simple redefinition of the parameters.) The fact that we found no statistically significant difference between the two approaches is consistent with their mathematical equivalence and supports our assertion that Brown's derivation of the Fokker-Planck equation [32] from the stochastic Langevin equation for the analogous problem in which the switching is induced by a magnetic field, has been successfully generalized to include current induced spin-torques.

**Effect of an initial Canting Angle**

Since we speculate that the exponentially decaying tail in the non-switched probability arises from the fact that the spin-torque vanishes at $\theta=0$, we investigated the effect of an initial canting angle on the switching probability.

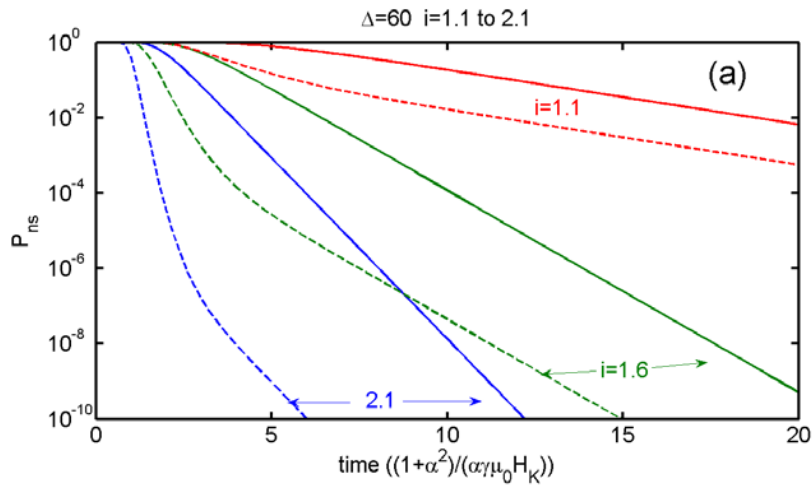

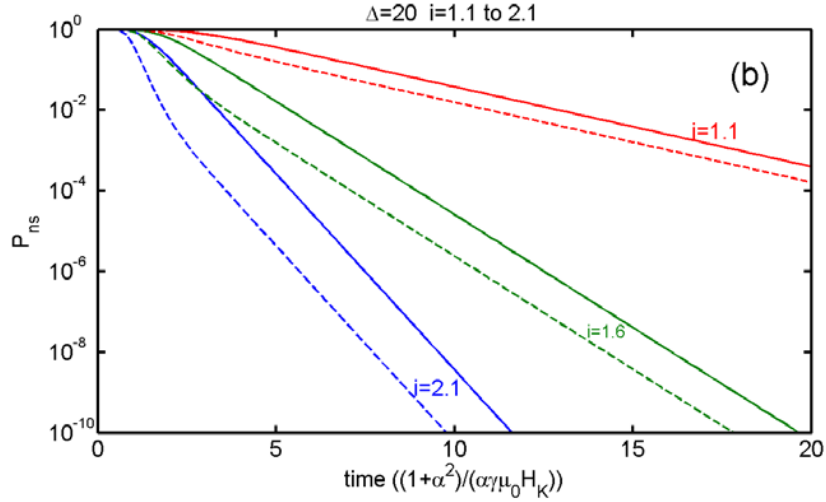

**Figure 6.** Dashed lines are the non-switched probability as a function of reduced time for the case that the initial angle is $\cos^{-1}(0.9) \sim 26°$. Solid lines are the numerical solutions to the Fokker-Planck equation for an initial thermal distribution. Panels (a) and (b) show the effect of an initial canting angle for thermal stability factors of 60 and 20 respectively.

Figure 6 shows the non-switched probability as a function of the reduced time for three values of the reduced current compared with the numerical solution to the Fokker-Planck equation for the case of an initial thermal distribution. From the figure, it can be seen that the non-switched fraction is significantly reduced if the initial distribution is canted, especially if $\Delta$ and $i-1$ are both large, however the exponential long-time tail persists with the same decay constant, $2\tau(i-1)$. For this example, the initial distribution had a maximum value of $\cos\theta$ of 0.9 and a minimum value of 0.89. For small values of $i-1$, the decrease in the $P_{ns}$ caused by an initial canting angle is smaller than for larger values, because there is a smaller spin-torque driving term so that the system spends more time at smaller angles allowing the diffusion term to establish a population at $\theta=0$. This is illustrated in Figure 6b which shows that an initial canting angle, has a much smaller effect for $\Delta=20$ compared to the case of $\Delta=60$ (Figure 6a) because of the larger diffusion term.

**Compare to Sun Switching Time Ansatz**

The exponential tail in the non-switched probability was anticipated by Sun et al. [46] and by He et al. [47] who postulated that the non-switched probability could be related to the initial thermal distribution through the relation (2.12) which approximately and deterministically relates the initial angle to the time to switch. Thus if one assumes that those (and only those) systems that have an initial angle greater than $\theta_0$ will have switched in time

$\tau_{sw} = \ln\left(\dfrac{\pi}{2\theta_0}\right)/(i-1)$, one can estimate the switched fraction at time $\tau$ as

$$P_{ns}(\tau) = \int_0^{\frac{\pi}{2}\exp[-\tau(i-1)]} \rho_0(\theta)\sin\theta\, d\theta. \qquad (4.5)$$

This approximation is shown in Figure 7 as the dashed curve labeled SST (Sun Switching Time Approximation). An improvement on this approximation can be made by using the exact expression for the switching time given in (2.14) which is shown in Figure 7 as the dashed curve marked CST (Corrected Switching Time Approximation),

$$P_{ns}\left(\tau_{sw}(x_0)\right) = \int_{x_0(\tau_{sw})}^{1} \rho_0(x)\, dx, \qquad (4.6)$$

where $\tau_{sw}(x_0)$ is given by (2.15). This latter curve is slightly greater than the approximate solution to the Fokker-Planck equation that we derived as Eq. (3.15) and show as the solid curve labeled AFP. The full solutions to the Fokker-Planck equation are given by the solid lines labeled FP. The small deviations from exponential decay that can be detected for the curves labeled SST and CST as $P_{ns}$ becomes small are due to integration errors. The asymptotic forms (for large values of $\tau$) for the SST and CST approximations for $P_{ns}$ are

$$P_{ns}^{SST} \to \frac{\Delta \pi^2}{4} \exp[-2\tau(i-1)], \qquad (4.7)$$

and

$$P_{ns}^{CST} \to \Delta 2^{\frac{2i}{i+1}} \left(1 - \frac{1}{i}\right)^{\frac{2}{i+1}} \exp[-2\tau(i-1)]. \qquad (4.8)$$

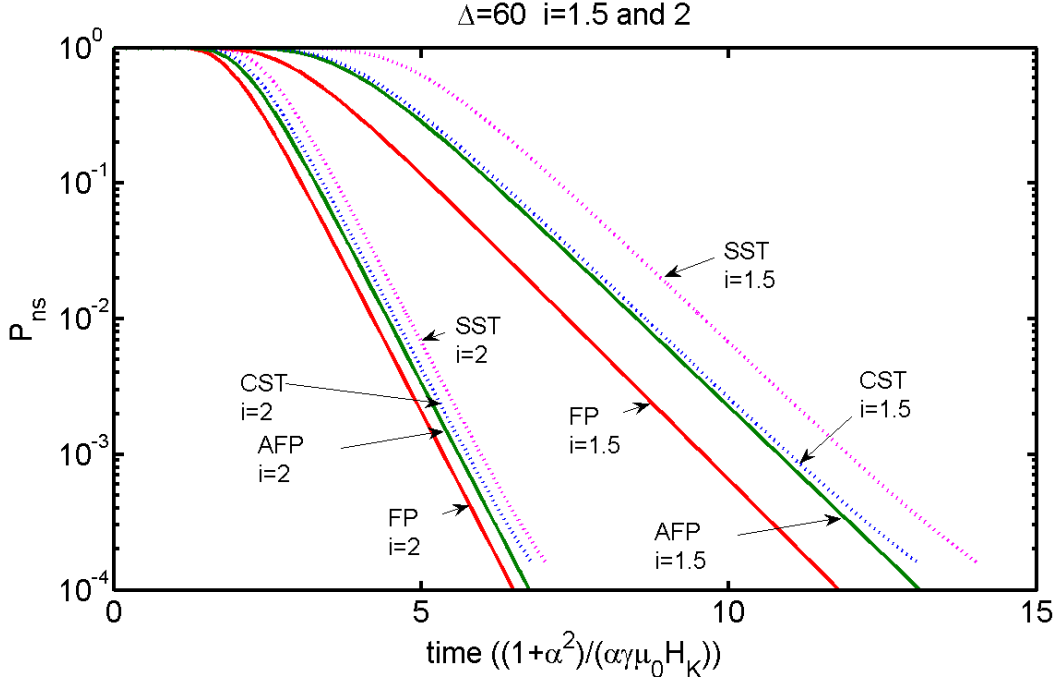

**Figure 7.** Comparison of four approximations for the time evolution of $P_{ns}(\tau)$, the probability that the element has not switched after reduced time, $\tau$. FP and AFP indicate respectively the numerical Fokker-Planck solution and the approximate analytical solution, based on (3.15). The curves SST and CST indicate respectively the estimation of the switching time from the initial probability density using the deterministic expressions for the switching time given by (2.12) and (2.15).

**Perturbing with in-plane components of pinned-layer magnetization or external field: stagnation points**

It might appear surprising that the STT approximation introduced above gives qualitatively correct results for the long time tail of the non-switching probability, because it is based on ignoring the thermal noise (deterministic motion) – basically the systems that don't switch are those that started sufficiently close to the point where there is no torque, which in our case is the film normal, $\theta = 0$ – in general, we will refer to this as a stagnation point. It seems that this involves a serious error, since a system that starts at the stagnation point will quickly be pulled away by random forces. We can understand why the approximation nevertheless works, by focusing on the sub-ensemble of our statistical ensemble which has a certain sequence of random fields over some particular time interval (but varying initial magnetization). These random fields produce a certain displacement $\Delta \mathbf{M}$ of the magnetization, which to lowest order is independent of the initial magnetization. Thus within this subensemble, there is still a stagnation point (from which the system ends up at $\theta = 0$ after this time interval) but the stagnation point is displaced by $-\Delta \mathbf{M}$. Thus the probability of ending up at $\theta = 0$ is the probability of having started at this displaced stagnation point, and if the noise is relatively small, this is the same as the probability of starting at the origin. In other words, the noise shifts the stagnation point, but does not eliminate it. In an initial thermal distribution, the stagnation point is unfortunately at the point of maximum probability. Fig. 6 showed that the WSER (probability of not switching) can be decreased by moving the initial distribution away from the stagnation point. However, it is not clear how this can be done physically. There is, however, a way of shifting the stagnation point away from the initial distribution, instead of *vice versa*. This can be done by tilting the external field, or by tilting the pinned-layer magnetization, which creates a spin torque even at $\theta = 0$. The total torque will vanish (we will have a stagnation point) only if this tilt torque is canceled by a precession torque, and this requires a tilt in $\mathbf{M}$. Thus the stagnation point will shift.

To do this quantitatively, we will generalize the linear approximation discussed in Sec. 4 above, by breaking the rotational symmetry that allowed us to work only with the polar angle $\theta$ (everything was independent of the azimuthal angle $\phi$). We will linearize about the normal direction (z), working only with the transverse component of the unit vector $\hat{\mathbf{m}}$, which we will denote by $\mathbf{m}$ for brevity:

$$\hat{\mathbf{m}} = \mathbf{m} + \cos\theta \hat{\mathbf{z}} \tag{4.9}$$

We may also replace $\cos\theta$ by 1.

We will also allow a tilted pinned magnetization direction, which we will write as

$$\mathbf{m}_p = \mathbf{m}_p^{\|} + m_{pz}\hat{\mathbf{z}} \tag{4.10}$$

although it will **not** be necessary to linearize in $\mathbf{m}_p^{\|}$ — the tilt angle can be as large as desired (although of course $\mathbf{m}_p^{\|}$ cannot exceed 1). We also introduce a dimensionless in-plane field $\mathbf{h}_{\|}$ defined by

$$\mathbf{H}_{\text{eff}} = H_K(1+h)\hat{\mathbf{z}} + H_K \mathbf{h}_{\|} \tag{4.11}$$

Linearizing in the in-plane component $\mathbf{m}$ gives us $\hat{\mathbf{m}} \times (\hat{\mathbf{m}} \times \hat{\mathbf{z}}) = \mathbf{m}$; linearizing also in the field tilt $\mathbf{h}_{\|}$ (dropping terms of order $\mathbf{m} \cdot \mathbf{h}_{\|}$) gives $\hat{\mathbf{m}} \times (\hat{\mathbf{m}} \times \mathbf{h}_{\|}) = -\mathbf{h}_{\|}$ so that the Landau-Lifshitz equation (1.3 – fix ref after merging) becomes

$$\begin{aligned}(1+\alpha^2)\frac{\partial \mathbf{M}}{\partial t} &= -\gamma\mu_0 M_s H_K(1+h)\mathbf{m}\times\hat{\mathbf{z}} - \alpha\gamma\mu_0 M_s H_K(1+h)\mathbf{m} - \alpha\gamma\mu_0 M_s H_K \mathbf{h}_{\|} \\ &\quad -\gamma\mu_0 M_s H_K \hat{\mathbf{z}}\times\mathbf{h}_{\|} - \frac{\gamma\hbar}{2eV}(\eta I - \alpha f)\hat{\mathbf{m}}\times(\hat{\mathbf{m}}\times\hat{\mathbf{m}}_p) + \frac{\gamma\hbar}{2eV}(f+\alpha\eta I)(\hat{\mathbf{m}}\times\hat{\mathbf{m}}_p)\end{aligned} \tag{4.12}$$

Note that in the absence of tilt and spin torque, the first term describes precession about the z axis with a frequency $\gamma\mu_0 H_K(1+h)/(1+\alpha^2)$ and the second describes dissipation at a rate $\alpha$ times this frequency (the spin torque term proportional to $I$ will modify the dissipation). The change of time variable to $\tau$, Eq. (2.6), removes this rate and leaves us with a dimensionless equation

$$\frac{\partial \hat{\mathbf{m}}}{\partial \tau} = -\alpha^{-1}(1+h)\mathbf{m}\times\hat{\mathbf{z}} - (1+h)\mathbf{m} - \mathbf{h}_\| - \alpha^{-1}\hat{\mathbf{z}}\times\mathbf{h}_\| \\ -\frac{1}{\alpha\gamma\mu_0 M_s H_K}\frac{\gamma\hbar}{2eV}(\eta I - \alpha f)\hat{\mathbf{m}}\times(\hat{\mathbf{m}}\times\hat{\mathbf{m}}_p) + \frac{1}{\alpha\gamma\mu_0 M_s H_K}\frac{\gamma\hbar}{2eV}(f+\alpha\eta I)(\hat{\mathbf{m}}\times\hat{\mathbf{m}}_p) \quad (4.13)$$

The coefficient of the triple product will be recognized as the dimensionless current $i$ defined in Eq. (ref) above:

$$i = \frac{1}{\alpha\gamma\mu_0 M_s H_K}\frac{\gamma\hbar}{2eV}(\eta I - \alpha f) \quad (4.14)$$

The last term is the field-like part of the current (with a small correction proportional to $\alpha I$), and we will denote its coefficient by

$$F = \frac{1}{\alpha\gamma\mu_0 M_s H_K}\frac{\gamma\hbar}{2eV}(f+\alpha\eta I) \quad (4.15)$$

Finally, we use $\hat{\mathbf{m}}\times\hat{\mathbf{m}}_p = \mathbf{m}\times\mathbf{m}_p^\| + m_{pz}\mathbf{m}\times\hat{\mathbf{z}} + \hat{\mathbf{z}}\times\mathbf{m}_p^\|$ and $\hat{\mathbf{m}}\times(\hat{\mathbf{m}}\times\hat{\mathbf{m}}_p) = m_{pz}\mathbf{m} - \mathbf{m}_p^\|$ and omit out-of-plane terms to obtain the dimensionless Landau-Lifshitz equation for the in-plane linearized magnetization:

$$\frac{\partial \mathbf{m}}{\partial \tau} = -\Omega \mathbf{m} \times \hat{\mathbf{z}} + \nu \mathbf{m} + \mathbf{T} \tag{4.16}$$

where

$$\Omega = \alpha^{-1}(1+h) - F \tag{4.17}$$

$$\nu = -1 - im_{pz} - h \tag{4.18}$$

(the sign of $\nu$ is chosen so that it is positive for switching -- note that in Sec. XX above, $m_{pz}$=-1) and all the azimuthal-symmetry-breaking effects are contained in the 2D "tilt" vector **T:**

$$\mathbf{T} = i\mathbf{m}_p^{\parallel} + F\hat{\mathbf{z}} \times \mathbf{m}_p^{\parallel} - \alpha^{-1}\mathbf{h}_{\parallel} \tag{4.19}$$

The Fokker-Planck equation for the linearized magnetization is

$$\frac{\partial \rho(\mathbf{m},\tau)}{\partial \tau} = -\nabla \cdot \mathbf{J}(\mathbf{m},\tau) \tag{4.20}$$

where the divergence is respect to the vector **m**, and the probability current is

$$\mathbf{J} = \rho \frac{d\mathbf{m}}{dt} - D\nabla \rho \tag{4.21}$$

To solve this equation, we generalize our ansatz (Sec. 3) to a Gaussian distribution with a drifting center $\mathbf{m}_d$ and a width parameter $W$:

$$\rho(\mathbf{m},\tau) = A(\tau)\exp\left[-\frac{(\mathbf{m}-\mathbf{m}_d(\tau))^2}{W(\tau)}\right] \tag{4.22}$$

Computing the necessary derivatives (with respect to **m**)

$$\nabla \rho = -\frac{2(\mathbf{m}-\mathbf{m}_d)}{W}\rho \tag{4.23}$$

$$\nabla^2 \rho = 4\frac{(\mathbf{m}-\mathbf{m}_d)^2 - W}{W^2}\rho \tag{4.24}$$

$$\nabla \cdot (\rho \mathbf{m}) = 2\frac{W - \mathbf{m}\cdot(\mathbf{m}-\mathbf{m}_d)}{W}\rho \tag{4.25}$$

$$\nabla \cdot (\rho \mathbf{m} \times \hat{\mathbf{z}}) = 2\frac{\rho}{W}(\hat{\mathbf{z}} \times \mathbf{m}_d) \cdot (\mathbf{m}-\mathbf{m}_d) \tag{4.26}$$

and substituting into the Landau-Lifshitz equation (Eq. (4.12)) gives an expression with various powers of (**m**–**m**$_d$). The coefficients of each power must match, giving for powers 0, 1, and 2:

$$\frac{dA}{d\tau} = -\left(2\nu + \frac{4D}{W}\right)A \tag{4.27}$$

$$\frac{d\mathbf{m}_d}{d\tau} = -\Omega \mathbf{m}_d \times \hat{\mathbf{z}} + \nu \mathbf{m}_d + \mathbf{T} \tag{4.28}$$

(exactly the deterministic part of the original LLG equation), and

$$\frac{dW}{d\tau} = 2\nu W + 4D \tag{4.29}$$

As in Sec. 3, we can relate D to the temperature by insisting that in equilibrium ($\nu$=-1, due to damping) the Boltzmann distribution ($W=1/\Delta$) is a time-independent solution:

$$D = \frac{1}{2\Delta} \tag{4.30}$$

It is convenient to make a distinction between the temperature governing the initial thermal distribution, whose corresponding stability factor (Eq. 2.8, Delta, fix when merge) will be denoted by $\Delta_0$, and the temperature governing the damping and random forces during switching, whose stability factor is $\Delta$. Then solving Eq. (4.29) with the initial condition $W(0)=1/\Delta_0$ gives an increasing width

$$W(\tau) = \frac{1}{\Delta_0} e^{2\nu t} + \frac{1}{\nu\Delta}\left(e^{2\nu t} - 1\right) \tag{4.31}$$

Normalization of the probability (Eq. (4.22)) requires that

$$A(\tau) = \frac{1}{\pi W(\tau)} \tag{4.32}$$

which is easily seen to be consistent with Eq. (4.27).

The drifting center spirals outward from a stagnation point

$$\mathbf{m}_s = -\frac{\Omega \mathbf{T} \times \hat{\mathbf{z}} + \nu \mathbf{T}}{\Omega^2 + \nu^2} \tag{4.33}$$

$$\mathbf{m}_d(\tau) = e^{\nu\tau}(\mathrm{Re}(Ae^{i\Omega\tau}), \mathrm{Im}(Ae^{i\Omega\tau}), 0) + \mathbf{m}_s \tag{4.34}$$

where the complex amplitude $A$ is determined by the initial condition:

$A = \left(\mathbf{m}_{dx}(0) - \mathbf{m}_{sx}\right) + \left(\mathbf{m}_{dy}(0) - \mathbf{m}_{sy}\right)i$, where $\mathbf{m}_d(0)=0$ usually. Our final exact result for the probability density is given by Eq. (4.22), together with Eqs. (4.32), (4.27), and (4.34).

To estimate the non-switching probability $P_{NS}(\tau)$, we choose an angle $\theta_{sw}$ (measured from the stagnation point, which is typically almost the same as measuring it from the origin) beyond which we will assume switching becomes inevitable (in Sec. 4 above, we took this to be $\pi/2$.) After most systems have switched (*i.e.*, in the long time tail of $P_{NS}$) we can neglect 1 in comparison with the exponential growth factor $e^{2\nu\tau}$, so the width becomes

$$W(\tau) \approx \left(\frac{1}{\Delta_0} + \frac{1}{\nu\Delta}\right)e^{2\nu t} \qquad (4.35)$$

Because this is much larger than the switching angle $\theta_{sw}$, the probability density is nearly constant over the switching circle, and we may approximate the integral by the value at $\mathbf{m}_s$ times the area $\pi\theta_{sw}^2$ of the circle. From Eq. (4.34), the $\left(\mathbf{m} - \mathbf{m}_d(\tau)\right)^2$ in Eq. (4.22) is $m_s^2\,e^{2\nu\tau}$, giving

$$P_{NS}(\tau) \approx \frac{\theta_{sw}^2}{\dfrac{1}{\Delta_0} + \dfrac{1}{\nu\Delta}}\, e^{-2\nu\tau}\, \exp\left[-\frac{m_s^2}{\dfrac{1}{\Delta_0} + \dfrac{1}{\nu\Delta}}\right] \qquad (4.36)$$

where (in the case of a pinned magnetization tilt $m_p^{\parallel}$) the stagnation point is given by

$$m_s^2 = \frac{T^2}{\left(\Omega^2 + \nu^2\right)} \approx \frac{\left(im_p^{\parallel}\right)^2}{\Omega^2} \qquad (4.37)$$

If we assume (as previously) that $\Delta = \Delta_0$, this becomes

$$P_{NS}(\tau) \approx \theta_{sw}^2 \Delta \frac{\nu}{\nu+1} e^{-2\nu\tau} \exp\left[-\frac{\nu}{\nu+1}\Delta m_s^2\right] \quad (4.38)$$

Note that this differs from our earlier analytic result for the symmetrical case (tilt $T=0$) only by the last exponential factor. We can relate this to a simple estimate of the effect of shifting the stagnation point away from the origin, that we gain a factor of the initial thermal probability of being at the stagnation point, the Boltzmann factor $\exp(-m_s^2 \Delta_0)$. In the limit of high overdrive $\nu$, when diffusion and noise is negligible, this is exactly correct; the additional term $1/\nu\Delta$ in Eq. (4.36) increases the nonswitching probability by taking into account diffusion back to the stagnation point of systems that don't start there. In general, as discussed previously in Section 4, we expect a linearized theory to work best for high overdrive. Fortunately, to get a reasonably low nonswitching probability it is normally necessary to use a fairly high overdrive.

As an additional test of the linearized theory, we can compare to our numerical results (above) for shifted initial conditions. Then the drifting center is initially at $\mathbf{m}_d(0)=(\theta_0,0,0)$ where $\theta_0 \sim 0.45$ is the initial angle. The initial spread of initial angle (proportional to $1/\Delta_0$) is zero. Then $\mathbf{m}_s = 0$, $A = \theta_0$, we again evaluate the probability (Eq. (4.22)) at the origin, where $(\mathbf{m}-\mathbf{m}_d(\tau))^2 = \mathbf{m}_d(\tau)^2 = e^{2\nu}\theta_0^2$. With Eq.(4.35), this gives

$$P_{NS}(\tau) \approx \theta_{sw}^2 \nu\Delta e^{-2\nu\tau} \exp\left[-\nu\Delta\theta_0^2\right] \quad (4.39)$$

The key practical question is, whether the stagnation point can in fact be shifted out (or mostly out) of the initial thermal probability distribution. Even if the tilt angle is large ($m_p^{\parallel} \sim 1$), and estimating ν ~ 1, i ~ 1, the dimensionless precession frequency Ω remains large (of order 1/α ~ 50 because of our time rescaling), so $m_s^2$ ~ 0.0004. Even if D is as high as 60, the gain factor is still negligible, exp(-0.024). We conclude that it is very difficult to shift the stagnation point by tilting the pinned magnetization. Imposing an in-plane magnetic field may be more promising, because of the factor 1/α in Eq. (4.19). A static bias field doesn't help, of course, since it will shift the initial distribution to the stagnation point and we lose the enhancement factor involving $m_s$ (the last exponential in eq. (4.36)). However, if we turn on the field when we turn on the current, Eq. (4.38) is valid with $m_s \approx h_\parallel$, and there is an improvement factor in $P_{NS}$ of

$$\exp\left[-\Delta \frac{\nu}{\nu+1} h_\parallel^2\right] \tag{4.40}$$

which can be substantial – if we set the tilt angle of the total field $h_\parallel$ to 0.45, with Δ=60 and ν=1 the factor becomes $e^{-6}$ ~ 0.0025.

**Soft Error Rates for Read Disturb**

Figure 8 shows the distribution function calculated for *Δ=60* and for *i=0.5* as a function of time. It is assumed that a current is applied for reading in order to determine whether the device is in a low (typically parallel moments of free and pinned layers) or a high (anti-parallel moments) resistance state. The applied current has a non-zero probability of causing or assisting a switching event. Spin-torque memory devices must be capable of switching from anti-parallel to parallel and from parallel to anti-parallel when information is being stored. Typically, the spin-torque efficiency is somewhat higher for anti-parallel to parallel switching than for parallel to anti-parallel [10, 44, 45] thus anti-parallel to parallel switching is usually easier. For reading one has the freedom to choose the current direction and thus one can use the current direction that stabilizes the anti-parallel configuration, i.e., electrons flow from free layer to pinned layer, thereby minimizing the probability of accidental switching into the parallel state.

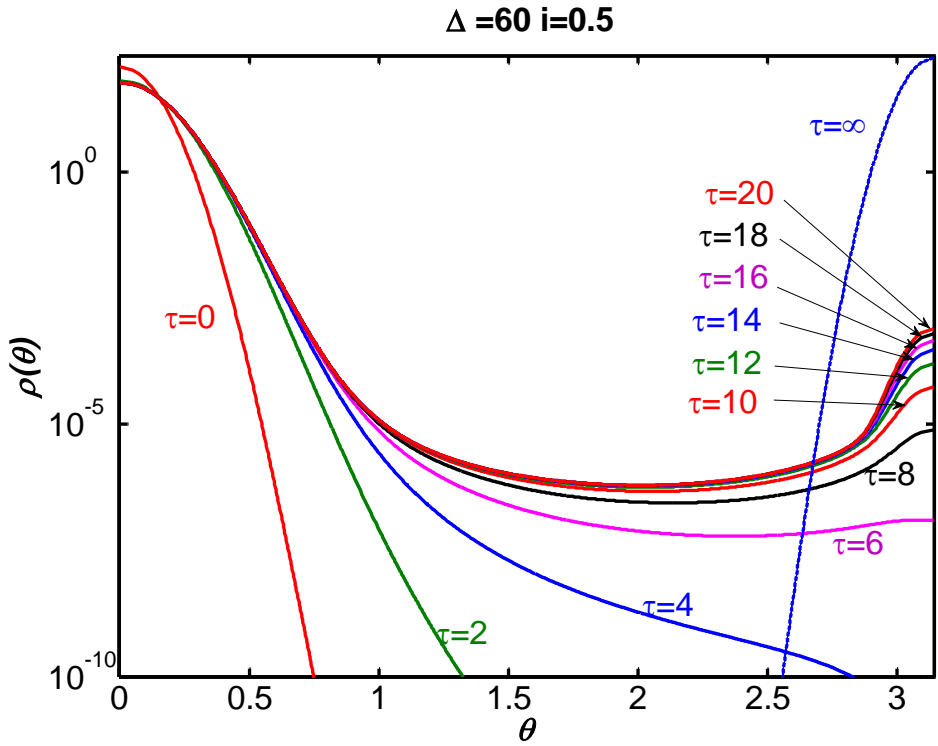

**Figure 8. Probability Distribution as a function of time for $\Delta$=60 and *i*=0.5.**

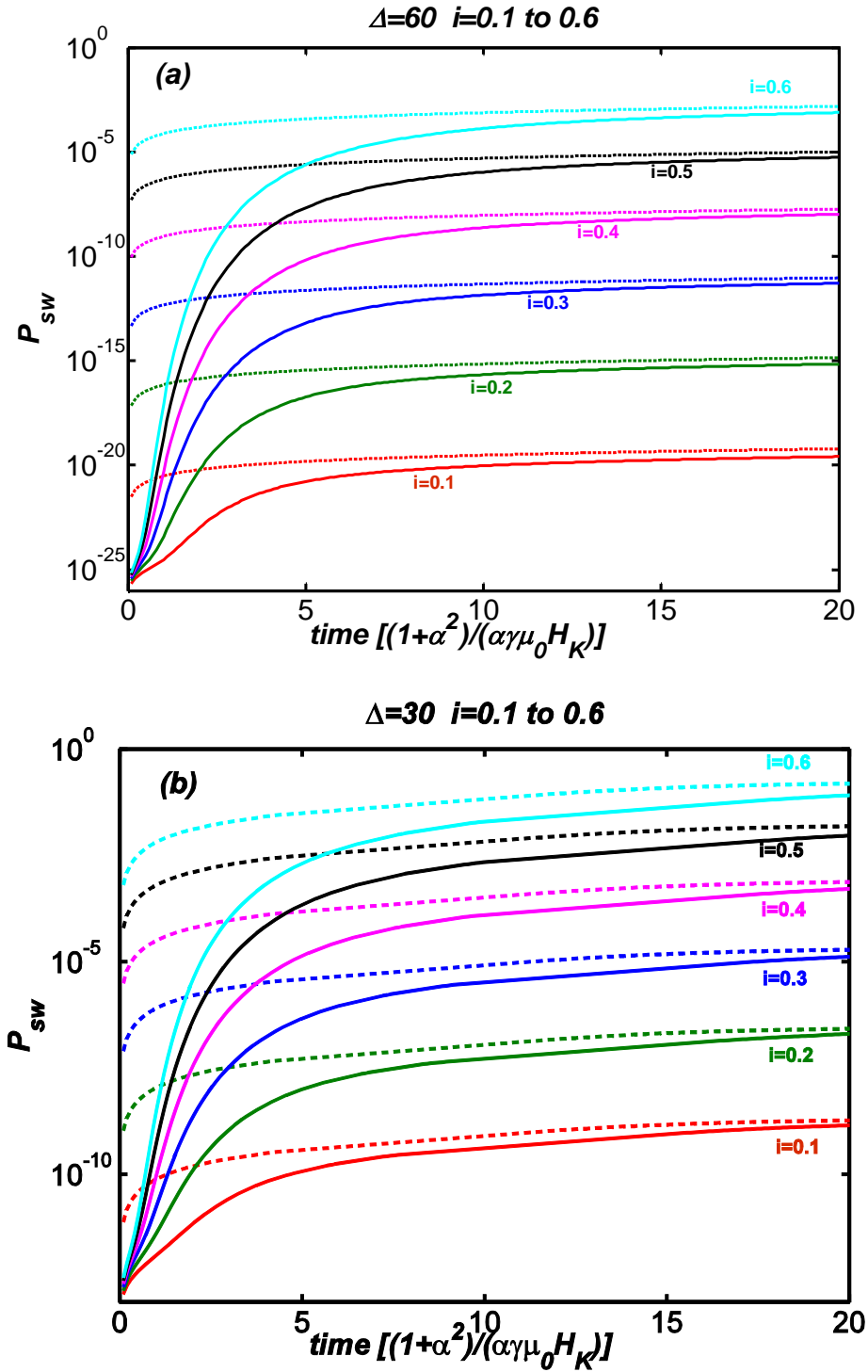

**Figure 9 RSER for Δ=60 and 30.**

Figure 9a shows the calculated read soft error rate (RSER) for Δ=60 and values of the reduced current between $i$=0.1 and $i$=0.6. The solid lines represent the numerical solutions to the Fokker-Planck equation. The dashed lines represent an approximate solution to the Fokker-Planck equation obtained by Brown for the case of a Stoner-Wohlfarth particle in an external magnetic field. We can apply his result to our case because of the correspondence between the spin-torque current and the applied magnetic field established in Equation (2.5).

The switching probability is observed to increase very rapidly initially, and then, after $\tau \approx 5$, the rate of increase slows and the switching probability is observed to increase linearly at a rate that increases with $i$. This can be understood in terms of a rapid equilibration within the well at $\theta = 0$ for $\tau <\sim 5$ followed by Brown-Kramers hopping over the energy barrier. The initial equilibration occurs because the effective energy function within the well changes when the current is applied. The Brown-Kramers approximation [32-34] to the switching probability is linear in $\tau$,

$$P_{sw} = \tau \sqrt{\frac{\Delta}{\pi}} \left(1-(i-h)^2\right) \left\{ (1-i+h)\exp\left[-\Delta(1-i+h)^2\right] + (1+i-h)\exp\left[-\Delta(1+i-h)^2\right] \right\} . (4.41)$$

The generalization of Brown's original derivation to include spin-torque is given in Appendix A. It can be seen from figures 9 and 10 that Equation (4.41) provides an upper limit for the switching probability and that it significantly overestimates the switching probability for reduced times less than ~ 10. Such times may be of interest for spin-torque devices, for example if $\alpha = 0.01$ and $\mu_0 H_K = 1T$, this would correspond to ~ 6 ns.

**Conclusion and Discussion**

In conclusion, we investigated spin-torque switching for devices in which the magnetization of the pinned and free layer is perpendicular to the plane of the layers in the macrospin approximation. Our investigation emphasized the time dependent probability for not switching when the applied current exceeds the critical current for switching and the probability for switching when the applied current is significantly below the critical current for switching. The former case determines the write soft error rate and the latter case determines the read soft error rate. Results are presented in terms of reduced currents and reduced time so that effectively all relevant cases are represented in the figures. We also provide approximate analytical formulas that can be used to estimate the read and write soft error rates. An important result is that the spin-polarized current enters the Fokker-Planck equation in essentially the same way as an axial magnetic field, allowing previous results derived for magnetic field induced switching to be used to describe current induced switching.

## Acknowledgments

This work was supported by the Defense Advanced Research Projects Agency's Office of Microsystems Technology and by Grandis, Inc. Useful conversations with D. Apalkov and E. Chen are gratefully acknowledged. We thank R. Goldfarb and M. Pufall for critical reviews of the manuscript and many helpful suggestions.

## Appendix

### Derivation of the generalized Brown-Kramers high energy barrier formula in the presence of spin-torque

In this appendix we present an approximate solution to the Fokker-Planck equation for the case in which the current is less than the critical current for switching. The approximation is valid in a limit in which the distribution function has come to quasi-equilibrium within each of two minima in the effective energy, but the total density in the two wells is not in equilibrium. The development follows closely Brown's [32] treatment of the analogous problem for field induced switching.

The Fokker-Planck equation is given by (3.8)

$$\frac{\partial \rho(\theta,\tau)}{\partial \tau} = -\nabla \cdot \mathbf{J}(\theta,\tau) \tag{A.1}$$

where,

$$J_\theta(\theta,\tau) = \left[\sin\theta(i-h-\cos\theta)\rho(\theta,\tau) - \frac{1}{2\Delta}\frac{\partial \rho(\theta,\tau)}{\partial \theta}\right]. \tag{A.2}$$

For very long times the system comes to equilibrium; $\frac{\partial \rho(\theta,\tau)}{\partial \tau} = 0$, and $J_\theta(\theta,\tau) = 0$. Using the latter result we have

$$\frac{1}{\rho(\theta,\infty)}\frac{\partial \rho(\theta,\infty)}{\partial \theta} = 2\Delta \sin\theta(i-h-\cos\theta).$$

Integrating from $\theta=0$ to $\theta$ yields

$$\ln \rho(\theta,\infty) - \ln \rho(0,\infty) = 2\Delta \int_0^\theta d\theta \sin\theta(i-h-\cos\theta) = -\Delta\left[\sin^2\theta + 2(i-h)(\cos\theta-1)\right] \tag{A.3}$$

or

$$\rho(\theta,\infty) = \rho(0,\infty)\exp\left\{-\Delta\left[\sin^2\theta + 2(i-h)(\cos\theta-1)\right]\right\}. \tag{A.4}$$

According to this result, the equilibrium distribution is appropriate to an effective energy function that has been modified from $E(\theta)/K_U^{\text{eff}} \equiv \varepsilon(\theta) = \sin^2\theta$ to $\varepsilon_{\text{eff}}(\theta) = \sin^2\theta + 2(i-h)(\cos\theta-1)$ by the spin polarized current. This effective energy function has minima at $\theta = 0$ and $\theta = \pi$ and a maximum at $\cos\theta = i\text{-}h$ if $i\text{-}h < 1$.

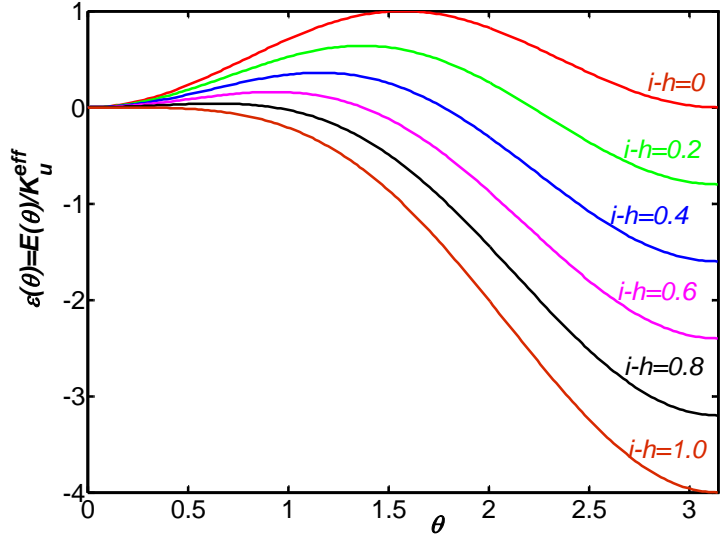

**Figure A.1 Effective energy function for values of *i-h* between 0 and 1.**

We now consider the case for which there is a maximum in the effective energy function (*i-h<1*) and the current pulse has been applied long enough for local equilibrium to be established in the vicinity of the bottom of the two wells, but there has not been sufficient time for the distribution to be equilibrated between the wells. If we assume that the effective energy function in the vicinity of each minimum is given by (A.4), we can relate the total probability of being in either well to the value of the distribution functions at the minima. For the well at $\theta = 0$, we have

$$n_0 = \rho_0 \int_0^{\theta_0} \sin\theta d\theta \exp\left\{-\Delta\left[\sin^2\theta + 2(i-h)(\cos\theta - 1)\right]\right\}, \quad (A.5)$$

and for the well at $\theta=\pi$,

$$n_\pi = \rho_\pi \int_{\pi-\theta_\pi}^{\pi} \sin\theta d\theta \exp\left\{-\Delta\left[\sin^2\theta + 2(i-h)(\cos\theta - 1) + 4(i-h)\right]\right\}. \quad (A.6)$$

Here, $\theta_0$ should be large enough and $\theta_\pi$ small enough that almost all of the distribution on the respective sides of the maximum are contained within the ranges $0 \leq \theta \leq \theta_0$ and $\theta_\pi \leq \theta \leq \pi$, respectively. We also require that $\theta_0$ be less than the angle for which the effective energy is a maximum and $\theta_\pi$ be greater ($\theta_0 < \cos^{-1}(i-h) < \theta_\pi$).

We can approximate the integrals in (A.5) and (A.6) by expanding the effective energy about the respective minima to order $\theta^2$, approximating $\sin\theta$ by $\theta$ and extending the range of integration to infinity, yielding

$$n_0 = \frac{\rho_0}{2\Delta(1-(i-h))} \qquad n_\pi = \frac{\rho_\pi}{2\Delta(1+(i-h))}. \tag{A.7}$$

Because equilibrium has not yet been established between the wells, there must be a current that flows between them. If we assume that there is no appreciable accumulation of the particles in the region between $\theta_0$ and $\theta_\pi$, we can calculate the current flowing between the wells. We concentrate on the interval $\theta_0 < \theta < \theta_\pi$. By assumption, $\frac{\partial \rho(\theta,\tau)}{\partial \tau} = 0$ for this region. This implies that $J_\theta(\theta,\tau)\sin\theta$ is constant in this region:

$$J_\theta(\theta,\tau)\sin\theta = \sin\theta\left[\sin\theta(i-h-\cos\theta)\rho(\theta,\tau) - \frac{1}{2\Delta}\frac{\partial \rho(\theta,\tau)}{\partial \theta}\right] = C. \tag{A.8}$$

This may be written as

$$\left[2\Delta\sin\theta(i-h-\cos\theta)\rho(\theta,\tau) - \frac{\partial \rho(\theta,\tau)}{\partial \theta}\right] = 2\Delta\frac{C}{\sin\theta}. \tag{A.9}$$

Multiplication of both sides by $\exp[\Delta\varepsilon_{\text{eff}}(\theta)]$ allows us to integrate the left-hand side between $\theta_0$ and $\theta_\pi$:

$$\frac{\partial}{\partial \theta}\left[\rho(\theta)\exp\left\{\Delta\left[\sin^2\theta + 2(i-h)(\cos\theta-1)\right]\right\}\right] = -2\Delta\frac{C}{\sin\theta}\exp\left\{\Delta\left[\sin^2\theta + 2(i-h)(\cos\theta-1)\right]\right\} \tag{A.10}$$

or

$$\left[\rho(\theta)\exp\left\{\Delta\left[\sin^2\theta+2(i-h)(\cos\theta-1)\right]\right\}\right]_{\theta_0}^{\theta_\pi} = -2\Delta C\int_{\theta_0}^{\theta_\pi}\frac{d\theta}{\sin\theta}\exp\left\{\Delta\left[\sin^2\theta+2(i-h)(\cos\theta-1)\right]\right\}$$

. (A.11)

Using the assumed variations of $\rho$ within the two wells, we have

$$\rho_\pi - \rho_0 = -2\Delta C\int_{\theta_0}^{\theta_\pi}\frac{d\theta}{\sin\theta}\exp\left\{\Delta\left[\sin^2\theta+2(i-h)(\cos\theta-1)\right]\right\}. \qquad (A.12)$$

The argument of the exponential has a maximum at $\cos\theta_m = (i-h)$. Expanding the argument about this maximum and approximating $\sin\theta$ by its value at the maximum, we have

$$\rho_\pi - \rho_0 = -2\Delta C\int_{\theta_0}^{\theta_\pi}\frac{d\theta}{\sin\theta}\exp\left\{\Delta\left[\sin^2\theta+2(i-h)(\cos\theta-1)\right]\right\}$$

$$\varepsilon_{\text{eff}}(\theta) = \sin^2\theta+2(i-h)(\cos\theta-1) \qquad \varepsilon_{\text{eff}}(\theta_m) = (i-h-1)^2$$
$$\varepsilon'_{\text{eff}}(\theta) = 2\sin\theta(\cos\theta-i+h) \qquad \varepsilon'_{\text{eff}}(\theta_m) = 0$$
$$\varepsilon''_{\text{eff}}(\theta) = 2\cos^2\theta-2\sin^2\theta-2(i-h)\cos\theta \qquad \varepsilon''_{\text{eff}}(\theta_m) = -2\left(1-(i-h)^2\right) \qquad (A.13)$$

$$\rho_\pi - \rho_0 = -\frac{2\Delta C\exp\left[\Delta(i-h-1)^2\right]}{\sqrt{1-(i-h)^2}}\int_{\theta_0}^{\theta_\pi}d\theta\exp\left[-\Delta\left(1-(i-h)^2\right)(\theta-\theta_m)^2\right]$$

$$C = \frac{(\rho_0-\rho_\pi)}{2\Delta}\sqrt{\frac{\Delta}{\pi}}\left(1-(i-h)^2\right)\exp\left[-\Delta(i-h-1)^2\right]$$

Using (A.7), we have

$$C = \left[n_0(1-i+h)-n_\pi(1+i-h)\right]\sqrt{\frac{\Delta}{\pi}}\left(1-(i-h)^2\right)\exp\left[-\Delta(i-h-1)^2\right] \qquad (A.14)$$

For the case where $n_\pi$ is still very small, the switched fraction will be given by

$$P_{sw} = \tau\sqrt{\frac{\Delta}{\pi}}(1-i+h))^2(1+i-h)\exp\left[-\Delta(1-i+h))^2\right]. \qquad (A.15)$$

This may be written as

$$\ln\left(\frac{P_{sw}}{\tau}\right) - \ln\left(\sqrt{\frac{\Delta}{\pi}}\right) = \ln\left[(1-i+h)^2(1+i-h)\right] - \Delta(1-i+h)^2 \qquad (A.16)$$